\documentclass{aa} 
\usepackage{txfonts}
\usepackage[latin1]{inputenc}  
\usepackage{graphicx}

\begin{document}
\title{Photodissociation of organic molecules\\
 in star-forming regions II: Acetic acid}
\author{S. Pilling\inst{1,2} \and A. C. F. Santos\inst{3} \and H. M. Boechat-Roberty\inst{1}}
\institute{Observatório do Valongo, Universidade Federal do Rio de
Janeiro, Ladeira Pedro Antônio 43, CEP 20080-090, Rio de Janeiro,
RJ, Brazil. \and Instituto de Química, Universidade Federal do Rio
de Janeiro, Ilha do Fundão, CEP 21949-900, Rio de Janeiro, RJ,
Brazil \and Instituto de Física, Universidade Federal do Rio de
Janeiro, Caixa Postal 68528, CEP 21941-972, Rio de Janeiro, RJ,
Brazil}
\offprints{S. Pilling,\\ \email{pilling@ov.ufrj.br}}
\date{Received / Accepted}
%
\abstract{Fragments from organic molecule dissociation (such as
reactive ions and radicals) can form interstellar complex molecules
like amino acids. The goal of this work is to experimentally study
photoionization and photodissociation processes of acetic acid
(CH$_3$COOH), a glycine (NH$_2$CH$_2$COOH) precursor molecule, by
soft X-ray photons. The measurements were taken at the Brazilian
Synchrotron Light Laboratory (LNLS), employing soft X-ray photons
from a toroidal grating monochromator (TGM) beamline (100 - 310 eV).
Mass spectra were obtained using the photoelectron photoion
coincidence (PEPICO) method. Kinetic energy distribution and
abundances for each ionic fragment have been obtained from the
analysis of the corresponding peak shapes in the mass spectra.
Absolute photoionization and photodissociation cross sections were
also determined. We have found, among the channels leading to
ionization, that only 4-6\% of CH$_3$COOH survive the strong
ionization field. CH$_3$CO$^+$, COOH$^+$ and CH$_3^+$ ions are the
main fragments, and the presence of the former may indicate that the
production-destruction process of acetic acid in hot molecular cores
(HMCs) could decrease the H$_2$O abundance since the net result of
this process converts H$_2$O into OH + H$^+$. The COOH$^+$ ion plays
an important role in ion-molecule reactions to form large
biomolecules like glycine.

}

\titlerunning{Photodissociation of organic molecules in SFRs II: Acetic acid}
\authorrunning{Pilling, Santos \& Boechat-Roberty}
\maketitle

\section{Introduction}

Acetic acid (CH$_3$COOH) has been detected toward various
astrophysical regions, including hot molecular cores (HMCs)
associated with low- and high-mass star-forming regions (Remijan
et al. 2002, 2003, 2004; Cazaux et al. 2003; Mehringer et al.
1997; Wotten et al. 1992 and Wlodarczak \& Demaiso 1988) and in
comets (Crovisier et al. 1999, 2004). In these environments the
radiation field can promote several photophysical and
photochemical processes, including the photodissociation. The
products of organic molecule dissociation (such as reactive ions
and radicals) can form interstellar complex molecules like long
carbon chain molecules and amino acids.

The simplest amino acid, glycine (NH$_2$CH$_2$COOH), was recently
detected in the molecular clouds SgrB2, Orion KL and W51 (Kuan et
al. 2003, 2004), although this result was questioned by Snyder et
al. (2005). In these objects, precursor molecules like ammonia
(NH$_3$), methylamine (CH$_3$NH$_2$), formic acid (HCOOH) and
acetic acid have been observed (Nummelin et al. 2000; Turner 1991
and Sutton et al. 1985). Liu et al. (2002) pointed out the
importance of performing studies on carboxyl acids since they
share common structural elements with biologically important
species such as amino acids.

Sgr B2, Orion KL and W51 are massive star-forming regions where
the presence of widespread UV and X-ray fields could trigger the
formation of photodissociation regions (PDRs). X-ray photons are
capable of traversing large column densities of gas before being
absorbed. X-ray-dominated regions (XDRs) in the interface between
the ionized gas and the self-shielded neutral layers could
influence the selective heating of the molecular gas. The
complexity of these regions possibly allows a combination of
different scenarios and excitation mechanisms to
coexist(Goicoechea et al. 2004).

The formation of interstellar acetic acid occurs both in the gas
phase and on grain surfaces. Huntress \& Mitchell (1979) proposed
a radiative association mechanism followed by dissociative
recombination with an electron in the gas-phase:
\begin{equation}
CH_3CO^+ + H_2O \longrightarrow CH_3COOH_2^+
\stackrel{e^-}{\longrightarrow} CH_3COOH + H.
\end{equation}
Ehrenfreund \& Charnley (2000) have proposed a warm gas-phase
route in which reactions with protonated methanol (CH$_3$OH$_2^+$)
and formic acid, evaporated from grain surfaces, can result in
CH$_3$COOH formation.

Sorrel (2001) has proposed that accretion of gas-phase H, O, OH,
H$_2$O, CH$_4$, NH$_3$ and CO onto dust grains sets up a
carbon-oxygen-nitrogen chemistry in the grain. As a consequence, a
high concentration of free OH, CH$_3$ and NH$_2$ radicals is
created in the grain mantle mainly by photolysis. Once these
radicals are created, they remain frozen in position until the
grains heat up. As this occurs, the radicals become mobile and
undergo chemical reactions amongst themselves and with other
adsorbed molecules to produce complex organic molecules including
CH$_3$COOH. Remijan et al. (2002) pointed out the importance of
grain-surface chemistry to form acetic acid, since the location of
acetic acid emission is nearly coincident with the emission from
other large molecular species formed by grain-surface chemistry in
HMCs.

Two isomers of acetic acid have also been detected toward HMC,
associated with star-forming regions, methyl formate (HCOOCH$_3$)
and glycolaldehyde (CH$_2$OHCHO). Hollis et al. (2001) have
measured the column density of both isomers in the Sgr B2 (LMH)
source near the galactic center and the relative abundances were
quite different, about 1:26:0.5
(CH$_3$COOH:HCOOCH$_3$:CH$_2$OHCHO). This suggests a different
pathway of formation and/or different stability against the
radiation field, which will be discussed in a future publication.

Remijan et al. (2004) have proposed that CH$_3$COOH formation
seems to favor HMCs with well-mixed N and O, despite the fact that
CH$_3$COOH does not contain an N atom. If this is proved to be
true, this is an important constraint on CH$_3$COOH formation and,
possibly, other structurally similar biomolecules. Despite the
differences in molecular structure and chemical formation
mechanisms, methyl cyanide, ethyl cyanide (CH$_3$CH$_2$CN) and
acetic acid are found to have similar abundances toward the W51 e1
and e2 sources.

The photodissociation of acetic acid has been studied
experimentally and theoretically in the ultraviolet region
(Bernstein et al. 2004; Maçôas et al. 2004, Fang et al. 2002; Naik
et al. 2001; Hunnicutt et al. 1989; Blake \& Jackson 1969).
However, despite some photoabsorption studies in the X-ray range
(Robin et al. 1998 and references therein) there are no studies
focusing on the photodissociation pathways due to soft X-rays. The
present work aims to examine the photoionization and
photodissociation of acetic acid by soft X-rays, from 100 eV up to
310 eV, including the energies around the carbon K edge ($\sim$
290 eV).

\section{Experimental setup}

The experiment was performed at the Brazilian Synchrotron Light
Laboratory (LNLS) in Campinas, S\~ao Paulo, Brazil. Soft X-rays
photons ($\sim10^{12}$ photons/s) from a toroidal grating
monochromator (TGM) beamline (100-310 eV), perpendicularly
intersect the gas sample inside a high vacuum chamber. The gas
needle was kept at ground potential. The emergent photon beam flux
was recorded by a light sensitive diode.
\begin{figure}[!htb]
\resizebox{\hsize}{!}{\includegraphics{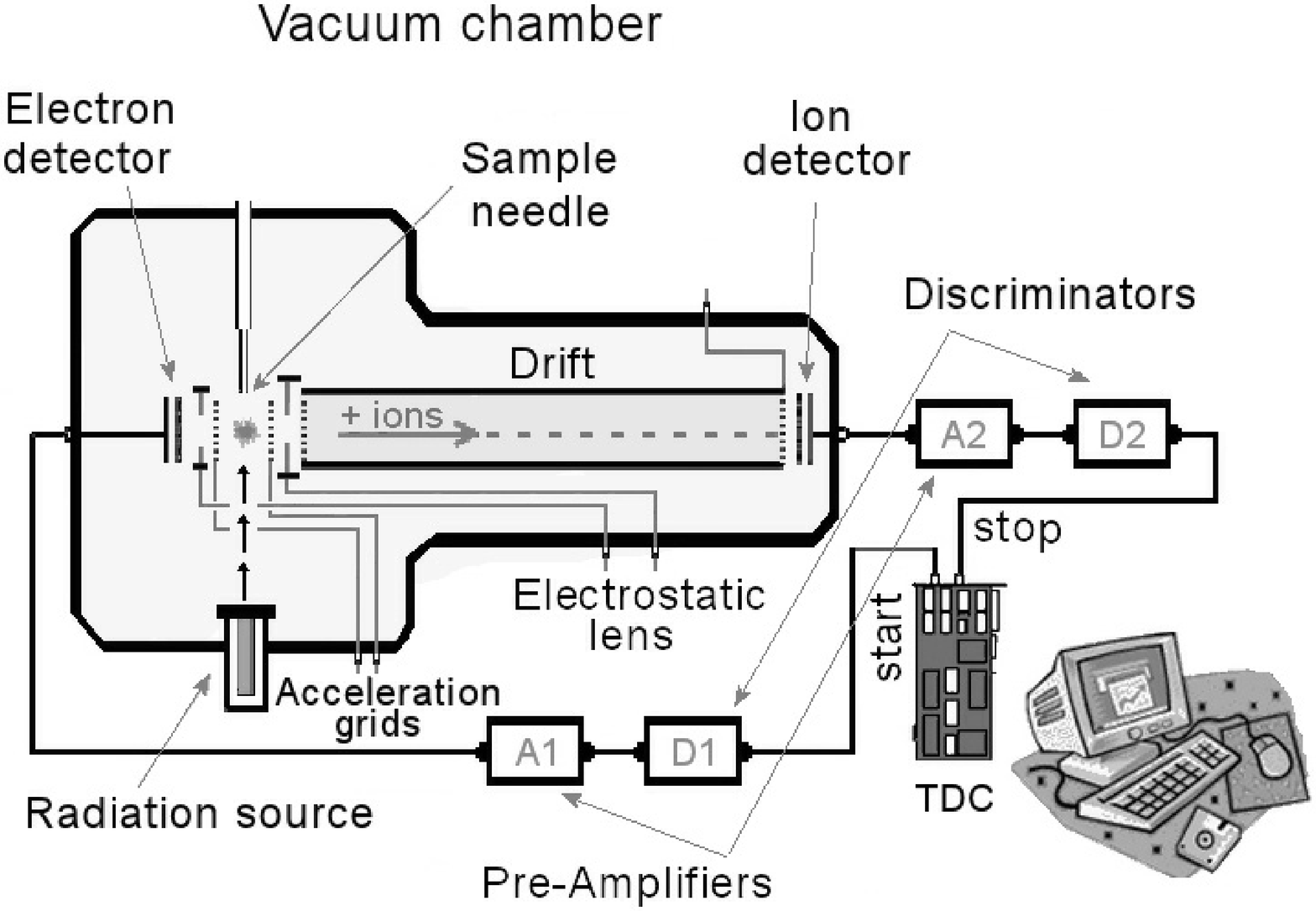}}
\caption{Schematic diagram of the time of flight mass spectrometer
inside the experimental vacuum chamber and the associated
electronics.} \label{fig-diagram}
\end{figure}
The sample was bought commercially from Sigma-Aldrich with purity
greate than 99.5\%. No further purification was performed other
than degassing the liquid sample by multiple freeze-pump-thaw
cycles before admitting the vapor into the chamber.

Conventional time-of-flight mass spectra (TOF-MS) were obtained
using the correlation between one Photoelectron and a Photoion
Coincidence (PEPICO). The ionized recoil fragments produced by the
interaction with the photon beam were accelerated by a two-stage
electric field and detected by two micro-channel plate detectors
in a chevron configuration, after mass-to-charge (m/q) analysis by
a time-of-flight mass spectrometer (297 mm long). They produced up
to three stop signals to a time-to-digital converter (TDC) started
by the signal from one of the electrons accelerated in the
opposite direction and recorded without energy analysis by two
micro-channel plate detectors. A schematic diagram of the time of
flight spectrometer inside the experimental vacuum chamber is
shown in Figure \ref{fig-diagram}, where A1 and A2 are the
pre-amplifiers and D1 and D2 are the discriminators. The
connection to the time-to-digital converter is also shown. Besides
PEPICO spectra, other two kinds of coincidence mass spectra were
obtained simultaneously, PE2PICO spectra (PhotoElectron Photoion
Photoion Coincidence) and PE3PICO spectra (PhotoElectron Photoion
Photoion Photoion Coincidence), which will be presented in a
future publication. These spectra have ions coming from double and
triple ionization processes, respectively, that arrive
coincidentally with photoelectrons. Of all signals received by the
detectors only about 10\% come from PE2PICO and 1\% from PE3PICO
spectra, reflecting that the majority contribution is indeed due
to single event coincidence. Nonetheless, PEPICO, PE2PICO and
PE3PICO signals were taken into account for normalization
purposes. Recoil ion and ejected electron detection efficiencies
of 0.23 and 0.04, respectively, were assumed. In addition, we
adopted the efficiencies of 0.54 and 0.78 to detect at least one
of the photoelectrons from double ionization and triple ionization
events, respectively (Cardoso 2001).

The first stage of the electric field (708 V/ cm) consists of a
plate-grid system crossed at the center by the photon beam . The
TOF-MS was designed to have a maximized efficiency for ions with
energies up to 30 eV. The secondary electrons produced in the
ionization region are focused by an electrostatic lens polarizing
the electron grid with 800~V, designed to focus electrons at the
center of the micro-channel plate detector. Negative ions may also
be produced and detected, but the corresponding cross-sections are
negligible.

The base pressure in the vacuum chamber was in the $10^{-8}$ Torr
range. During the experiment the chamber pressure was maintained
below $10^{-5}$ Torr. The pressure at the interaction region
(volume defined by the gas beam and the photon beam intersection)
was estimated to be $\sim$ 1 Torr (10$^{16}$ mols cm$-3$). The
measurements were made at room temperature.

\section{Results and discussion}

\begin{figure}[!htb]
\resizebox{\hsize}{!}{\includegraphics{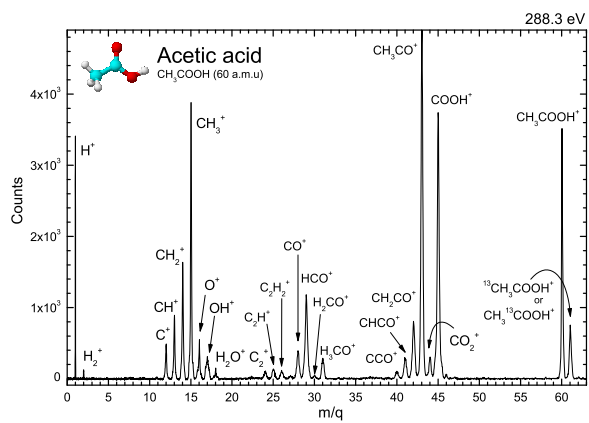}}
\caption{Time-of-flight mass spectra of acetic acid molecule
recorded at 288.3 eV.} \label{fig-ms}
\end{figure}

Figure~\ref{fig-ms} shows the mass spectrum of acetic acid
obtained at photon energy of 288.3 eV. The
C1s$\rightarrow\pi^{\ast}$ resonance energy from C=O bond is 288.6
eV (Robin et al 1988). We can see the methyl fragment group (mass
from 12 to 15 a.m.u), the C$_2$ group (24 to 27 a.m.u), the HCO
group (28 to 31 a.m.u) and the CCO group (40 to 44 a.m.u). The
carbon dioxide  ion (CO$_2^+$) is also produced. At mass 45 a.m.u.
we see the carboxyl ion (COOH$^+$). $^{13}$CH$_3$COOH$^+$ (or
CH$_3^{13}$COOH$^+$) is also seen.

The ions with the largest yields are the ionized acetyl radical
(CH$_3$CO$^+$), ionized methyl (CH$_3^+$) and the ionized carboxyl
radical (COOH$^+$). The first two may be interpreted as a result
of neutral OH and COOH liberation, due to the bond rupture near
the carbonyl in the dissociation process.

\begin{figure}[h]
\resizebox{\hsize}{!}{\includegraphics{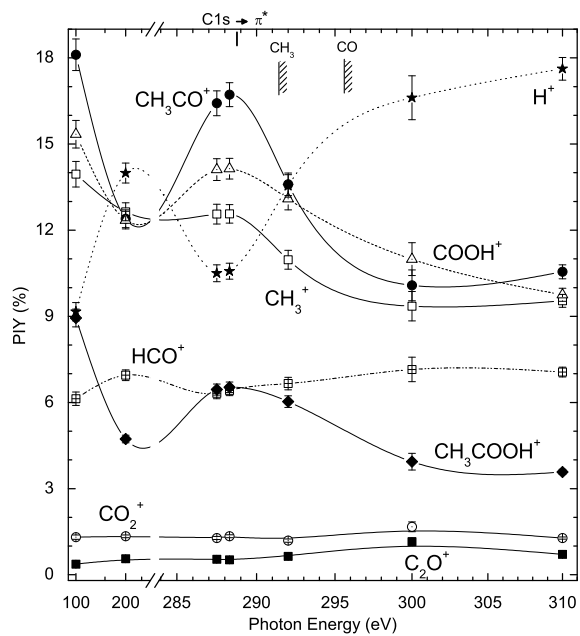}}
\caption{Partial ion yield (PIY) of some PEPICO fragments of
CH$_3$COOH molecule as a function of photon energy.} \label{fig-piy}
\end{figure}

The CH$_3^+$ and COOH$^+$ fragments present similar relative
partial yield which may indicate that the molecular dissociation
due to photoionization of the C1s inner shell-electron at 288.3 eV
does not have a preferential carbon atom target. In this case,
both molecular fragments could retain the core hole (charge).
However, as we can see in Figure~\ref{fig-piy} and in
Table~\ref{tab-piy}, there was a small enhancement ($\leq$ 20\%)
in the production of COOH$^+$ with respect to CH$_3^+$ in the
photon energy range of 288 to 300 eV. This behavior may be
associated with some instabilities of CH$_3^+$ ion leading to
subsequence fragmentation and/or charge resonance in the COOH$^+$
ion, increasing its stability.


Figure~\ref{fig-piy} shows the partial ion yields (PIY) for the
most significant outcomes (CH$_3$CO$^+$, COOH$^+$, H$^+$ and
CH$_3^+$) in the acetic acid dissociation in 100-310 eV photon
energy range. The yields of some minor products like HCO$^+$,
CH$_3$COOH$^+$, CO$_2^+$ and C$_2$O$^+$ are also shown. A clear
bump can be seen in the fractions of CH$_3$CO$^+$, COOH$^+$,
CH$_3^+$ and CH$_3$COOH$^+$ near the C1s resonance and decreasing
to higher photon energies. The  HCO$^+$ and H$^+$ fragments show a
distinct behavior and also exhibit a gradual increase toward
higher energies which indicates that these ions are preferentially
formed after the normal Auger decay. The C$_2$O$^+$ (0.9\%) and
CO$_2^+$ (1.3\%) fragments do not show any clear energy
dependence. The C1s resonance and the ionization potential of each
carbon are also indicated. The statistical uncertainties are below
10\%.

\begin{figure}[h]
\resizebox{\hsize}{!}{\includegraphics{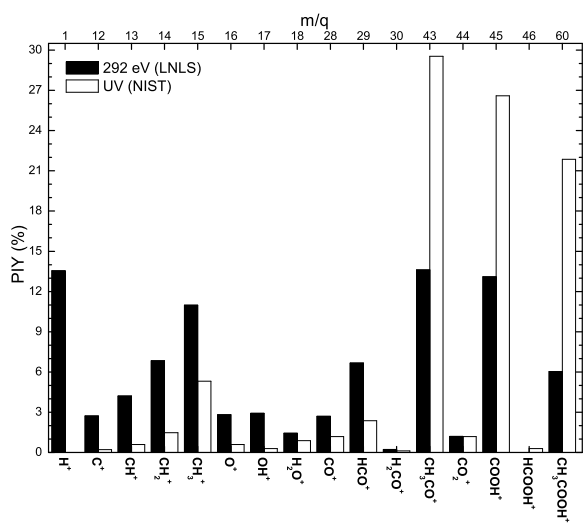}}
\caption{Comparison of partial ion yield (PIY) of acetic acid
fragments in soft X-ray and UV radiation.}
 \label{fig-piycomp}
\end{figure}

In Figure \ref{fig-piycomp} we compared the partial ion yield in
soft X-rays (292 eV) obtained at LNLS and UV (by 70 eV
electrons)\footnote {The effect of 70 eV electrons is very similar
to 21.21 eV (He I Lamp) photons in both cases the main ionization
occurs in valence shells (see discussion in Lago et al. 2004)}
from NIST\footnote {National Institute of Standards and Technology
http://webbook.nist.gov/chemistry/}. The molecular ion
CH$_3$COOH$^+$ is more destroyed by soft X-rays than by UV
photons, as expected. The partial ion yields of several fragments
are also different in X-ray and in UV fields, for example, the
enhancement of COOH$^+$ and CH$_3$CO$^+$ produced by UV radiation.
The opposite occurs with HCO$^+$ and all lower mass ions, which
seem to be more efficiently produced by X-ray photons.

Practically no production of HCOOH$^+$ (formic acid ion) has been
identified in the fragmentation by soft X-rays and, while UV
photons produce only a small amount of this ion, as we can see in
Figure~\ref{fig-piy} and \ref{fig-piycomp}. As was noted before,
the CO$_2^+$ and the C$_2$O$^+$ yields remain unchanged in the
UV-X-ray photon energy range. The same applies to H$_2$CO$^+$
(formaldehyde) despite the extremely low yield (Figure
\ref{fig-piycomp}).

The largest production of ions due to the photodissociation by
soft X-ray photons near the C1s resonance energy is for
CH$_3$CO$^+$, followed by COOH$^+$, CH$_3^+$ and H$^+$. The
former, as mentioned before, was suggested by Huntress \& Mitchell
(1979) to be the main precursor of acetic acid together with the
water molecule. Therefore, the continuous formation-destruction
cycle of acetic acid could act as a catalytic agent converting
water molecules in OH + H or (O + 2H) and possibly promoting a
reduction in the local H$_2$O abundance. For higher photon
energies, as we can see in Figure~\ref{fig-piy}, the proton
production is the major outcome.

The carboxyl radical, as has been discussed by several authors
(Woon 2002, Largo et al. 2004 and Mendonza et al. 2004), plays an
important role in the formation of large biomolecules including
amino acids. Its ionic counterpart was one of the most ionic
products released from the photodissociation of acetic acid by a
moderate/strong ionizing photon field. This may suggest that never
though acetic acid undergoes considerable photodissociation in
star-forming region, it could still produce glycine and other
carboxylated biomolecules via barrier-free ion-molecule reactions
involving some of its reactant fragments like COOH$^+$.

\subsection{Kinetic energy release (heating) of the ionic fragments}

Several authors have recently focused on the pathway of formation
of biomolecules present in the star-forming region and other
gaseous-dusty astronomical media (Largo 2004, Woon 2002, and
references therein). Despite the success of ab initio theoretical
calculations, the endothermic ion-molecule reactions have been
neglected and only exothermic reactions have been accepted as a
viable mechanism. However, with the knowledge of the kinetic
energy (or at least with its value range) of some radical and
ionic fragments, some endothermic ion-molecule reactions could be
likely and, in extreme situations, or even become more efficient
than exothermic reactions.

We have determined the kinetic energy of all cationic fragments from
the photodissociation of acetic acid. The present time-of flight
spectrometer was designed to fulfil the Wiley-McLaren conditions for
space focusing (Wiley \& McLaren 1955). Within the space focusing
conditions, the observed broadening of peaks in spectra is mainly
due to kinetic energy release of fragments. Considering that the
electric field in the interaction region is uniform, we can
determine the released energy in the fragmentation process ($U_0$)
from each peak width used by Simon et al. (1991), Hansen et al.
(1998) and Santos, Lucas \& de Souza (2001)

\begin{equation} \label{eq-U0}
U_0 = \Big(\frac{qE \Delta t}{2} \Big)^2 \frac{1}{2m}
\end{equation}
where $q$ is the ion fragment charge, $E$ the electric field in the
interaction region, $m$ is the mass of the fragment, and $\Delta t$
is the time peak width (FWHM) taken from PEPICO spectra. In order to
test the above equation we have measured the argon mass spectrum
under the same conditions.

The calculated values for kinetic energy release ($U_0$) for acetic
acid fragmentation are shown in Table~\ref{tab-piy}. We observe that
the highest kinetic energy release was associated with the lightest
fragment H$^+$ ($m/q=1$) followed by $H_2^+$ ($m/q=2$), as expected.
Differently from formic acid photodissociation results
(Boechat-Roberty et al. 2005), extremely fast ionic fragments ($U_0>
10$ eV), usually associated with dissociation of doubly or
multiply-charged ions, were not observed at high photon energies.

The study of the decay of core-excited molecules provides
information about the bonding or antibonding nature of the
molecular orbitals. Generally, the final electronic states of a
core excited molecule are unknown due to the fact that the
densities of the states are very high, and the bond distances and
angles differ from their ground state configuration. The surface
potentials of the ionic states are extremely repulsive. For core
excited molecules that dissociate into one charged and one or more
neutral fragments, the dissociation is primarily controlled by
chemical (non-Coulomb) forces originating from the residual
valence electrons of the system (Nenner \& Morin, 1996). From
Table 1, one can see that the mean kinetic energy release $U_0$ of
some acetic acid fragments increases as the photon energy
approaches the C 1s edge (288 eV). This enhancement is due to the
repulsive character of the $\sigma$* ($\pi$*) resonance.

\begin{table*}
\centering \caption{Relative intensities (Partial Ion Yield - PIY)
and kinetic energy $U_0$ release by fragments in the acetic acid
mass spectra, as a function of photon energy. Only fragments with
intensity $>$ 0.1 \% were tabulated. The estimated experimental
error was below 10\%.} \label{tab-piy}
\begin{tabular}{ l l l r r r r r r r }
\hline \hline
\multicolumn{2}{c}{Fragments}    &  & \multicolumn{7}{c}{PIY (\%) / $U_0$ (eV)}\\
\cline{1-2}  \cline{4-10}
   $m/q$        & Attribution    &  & 100 eV       & 200 eV       & 287.5 eV      & 288.3 eV    & 292 eV       & 300 eV       & 310 eV       \\
\hline
1       & $H^+$                  &  & 9.16 / 2.9   & 14.0 / 2.9   & 10.5 / 2.9   & 10.6 / 2.9   & 13.5 / 3.8   & 16.6 / 3.8   & 17.6 / 4.9  \\
2       & $H_2^+$                &  & -            & 0.59 / 3.0   & 0.53 / 2.4   & 0.41 / 1.5   & 0.58 / 3.6   & 0.77 / 1.9   & 0.63 / 3.0  \\
12      & $C^+$                  &  & 1.98 / 0.24  & 2.69 / 0.31  & 2.10 / 0.31  & 2.08 / 0.24  & 2.71 / 0.40  & 3.71 / 0.32  & 3.66 / 0.50  \\
13      & $CH^+$                 &  & 3.48 / 0.22  & 4.47 / 0.22  & 3.81 / 0.22  & 3.54 / 0.29  & 4.20 / 0.29  & 5.24 / 0.37  & 4.87 / 0.29  \\
14      & $CH_2^+$; $CO^{++}$    &  & 6.71 / 0.21  & 7.38 / 0.21  & 6.60 / 0.21  & 6.54 / 0.20  & 6.84 / 0.27  & 7.71 / 0.20  & 7.61 / 0.34 \\
15      & $CH_3^+$               &  & 13.9 / 0.14  & 12.6 / 0.19  & 12.6 / 0.10  & 12.6 / 0.14  & 10.9 / 0.09  & 9.35 / 0.14  & 9.54 / 0.14  \\
16      & $O^+$                  &  & 2.19 / 0.13  & 2.88 / 0.37  & 2.31 / 0.13  & 2.29 / 0.23  & 2.81 / 0.13  & 3.42 / 0.73  & 3.28 / 0.63 \\
17      & $OH^+$                 &  & 2.55 / 0.68  & 2.81 / 1.14  & 2.34 / 0.79  & 2.21 / 1.1   & 2.91 / 0.35  & 3.26 / 0.69  & 2.89 / 1.86 \\
18      & $H_2O^+$               &  & 0.61 / 0.27  & 0.76 / 0.12  & 0.69 / 0.08  & 0.56 / 0.05  & 1.43 / 0.05  & 1.47 / 0.53  & 0.60 / 0.11  \\
24      & $C_2^+$                &  & -            & 0.88 / 0.56  & 0.63 / 0.35  & 0.58 / 0.48  & 1.03 / 0.42  & 1.50 / 0.25  & 1.17 / 0.56  \\
25      & $CHC^{+}$ ?            &  & -            & 1.14 / 0.40  & 0.89 / 0.40  & 0.80 / 0.40  & 1.07 / 0.40  & 1.50 / 0.95  & 1.17 / 0.54  \\
26      & $CH_2C^{+}$ ?          &  & -            & 0.87 / 0.27  & 0.68 / 0.23  & 0.65 / 0.45  & 0.84 / 0.33  & -            & 0.77 / 0.58  \\
28      & $CO^+$                 &  & 1.98 / 0.14  & 2.72 / 0.30  & 2.26 / 0.26  & 2.13 / 0.26  & 2.69 / 0.26  & 3.96 / 0.10  & 3.36 / 0.42 \\
29      & $COH^+$; $HCO^+$       &  & 6.13 / 0.16  & 6.95 / 0.25  & 6.32 / 0.21  & 6.42 / 0.21  & 6.66 / 0.29  & 7.15 / 0.17  & 7.06 / 0.40 \\
30      & $CH_2O^+$; $CH3COOH^{++}$ ? & & 1.33 / 0.23 & -          & 0.27 / 0.24  & 0.55 / 0.05  & 0.22 / 0.45  & -          & 0.21 / 0.16 \\
31      & $CH_3O^{+}$ ?          &  & -            & 1.27 / 0.23  & 1.42 / 0.33  & 1.41 / 0.27  & 1.29 / 0.23  & 1.43 / 0.07  & 0.92 / 0.27  \\
40      & $CCO^+$                &  & 0.37 / 0.01  & 0.56 / 0.18  & 0.54 / 0.21  & 0.52 / 0.22  & 0.64 / 0.15  & 1.15 / 0.18  & 0.71 / 0.18 \\
41      & $CHCO^+$               &  & 1.17 / 0.12  & 1.67 / 0.12  & 1.41 / 0.14  & 1.39 / 0.18  & 1.49 / 0.12  & 1.78 / 0.05  & 1.59 / 0.24 \\
42      & $CH_2CO^+$             &  & 3.47 / 0.09  & 3.69 / 0.14  & 3.70 / 0.14  & 3.63 / 0.11  & 3.56 / 0.14  & 3.21 / 0.07  & 3.31 / 0.17 \\
43      & $CH_3CO^+$;            &  & 18.1 / 0.06  & 12.4 / 0.07  & 16.4 / 0.07  & 16.7 / 0.07  & 13.6 / 0.07  & 10.1 / 0.05  & 10.6 / 0.06 \\
44      & $CO_2^+$;$CH_3COH^+$   &  & 1.31 / 0.03  & 1.34 / 0.07  & 1.28 / 0.11  & 1.34 / 0.11  & 1.19 / 0.09  & 1.67 / 0.14  & 1.28 / 0.11  \\
45      & $COOH^+$               &  & 15.3 / 0.06  & 12.4 / 0.08  & 14.1 / 0.08  & 14.1 / 0.08  & 13.1 / 0.05  & 11.0 / 0.11  & 9.75 / 0.08  \\
46      & $HCOOH^+$              &  & -            & -            & 0.15 / 0.06  & 0.16 / 0.03  & -            & -            & 0.12 / 0.06  \\
60      & $CH_3COOH^+$           &  & 8.94 / 0.01  & 4.73 / 0.01  & 6.45 / 0.01  & 6.53 / 0.01  & 6.03 / 0.01  & 3.94 / 0.02  & 3.58 / 0.02  \\
61      & $^{13}CH_3COOH^+$      &  & 1.13 / 0.02  & 1.16 / 0.06  & 1.71 / 0.03  & 2.26 / 0.03  & 0.21 / 0.02  & -            & 2.45 / 0.05  \\
\hline \hline
\end{tabular}
\end{table*}

\subsection{Photodissociation and formation pathways}

Acetic acid is one of the simplest carboxylic acids. Its
underlying decomposition dynamics (dissociation pathways) have
been extensively investigated in the UV range, from both
experimental and theoretical points of view. On the basis of the
stable products observed in UV photolysis gas-phase of acetic
acid, several possible dissociation processes (Fang et al. 2002;
Mackie \& Doolan 1984; Satio et al. 1990; Blake \& Jackson 1969;
Ausloss \& Steacie 1955 ) were suggested:
\\
\begin{eqnarray}
CH_3COOH + h\nu       &\longrightarrow& CH_3CO + OH \\
                      &\longrightarrow& CH_3 + COOH \\
                      &\longrightarrow& CH_3CO_2 + H \\
                      &\longrightarrow& CH_2COOH + H \\
                      &\longrightarrow& CH_4 + CO_2 \\
                      &\longrightarrow& CH_2CO + H_2O
\end{eqnarray}
This was also confirmed by Hunnicutt et al. (1989) using both
room-temperature and jet-cooled conditions. The major pathway for
acetic acid UV photodissociation is reaction 3. Moreover, the
observed isotropic distribution of OH fragments from acetic acid
dissociation and the parent molecule fluorescence is indicative of
a moderately slow dissociation. Reactions 4 and 5 could also lead
to CH$_3$ + CO + OH products (Hunnicutt et al. 1989).

The behavior of acetic acid in the ice phase under the influence
of UV photons was recently studied by Maçôas et al. (2004) in an
experimental/theoretical approach. The UV photolysis of the Ar
matrix-isolated acetic acid reveals very different products from
the gas phase. As a result, 37\% of the acetic acid molecules
yield methanol (CH$_3$OH) plus carbon monoxide complexes, 17\%
yield carbon monoxide complexed with formaldehyde and molecular
hydrogen, 20\% yield quaternary complexes of two carbon monoxides
molecules and two hydrogen molecules and 21\% dissociate into
carbon dioxide and methane (CH$_4$). Ketene (CH$_2$CO), which is
the main product of thermal decomposition, was detected only in
small amounts ($\leq$ 5\%). The CO/CO$_2$ product ratio is $\sim$
5.

Bernstein et al. (2004) also studied acetic acid in the ice-phase,
using both Ar and water matrices and determined the half-life in a
UV radiation field and revealed a relatively low survival of
acetic acid (large yield of photoproducts) when compared to other
organic molecules, like aminoacetonitrile (H$_2$NCH$_2$CN). These
experiments  demonstrate that organic nitriles (cyanide compounds)
survive 5 to 10 times longer exposure to UV photolysis than do the
corresponding acids.

The present work shows significative differences between the
photoproducts of acetic acid due to low ionizing radiation (UV)
and soft X-ray photons. The inner shell photoionization process
may produce instabilities in the molecular structure (nuclear
rearrangements) leading to dissociation. From Table~\ref{tab-piy}
we determine the main photodissociation pathways from single
ionizations in 100-300 eV photon energy range. These
photodissociation pathways are shown in Table~\ref{tab-path1}.
Only events with greater than 1\% yields were considered here. The
main dissociation leads to production of CH$_3$CO$^+$ due to the
bond rupture of OH near the carbonyl.

Herbst \& Leung (1986) have presented several pathway syntheses of
complex molecules in dense interstellar clouds via gas-phase
chemistry models. The authors presented a significant amount of
normal ion-molecule reactions including the ions C$^+$, C$_2^+$,
CH$^+$, OH$^+$, CO$^+$, CH$_2^+$, H$_2$O$^+$, HCO$^+$, C$_2$H$^+$,
CO$_2^+$, CH$_3^+$. In another set of reactions they have shown
several radiative associations including the ions C$^+$, CH$_3^+$,
HCO$^+$ leading to the production of high molecular complexity
species. For example, the possible route to interstellar
acetaldehyde, involving the HCO$^+$ ion and methane by
\begin{equation}
HCO^+ + CH_4 \longrightarrow C_2H_5O^+ + h\nu
\stackrel{e^-}{\longrightarrow} CH_3CHO + H.
\end{equation}

Combes et al. (1987) have suggested that the gas phase acetone
((CH$_3$)$_2$CO) present in the Sgr B2 region could be formed by
the radiative association reaction between the methyl ion and
interstellar acetaldehyde, followed by dissociative recombination
with electrons by
\begin{equation}
CH_3CHO + CH_3^+ \rightarrow CH_3COCH_4^+ + h\nu
\stackrel{e^-}{\rightarrow} (CH_3)_2CO + H.
\end{equation}

Other radiative association reactions with the methyl ion could
produce the simplest interstellar ether. In this situation, as
proposed by Herbst \& Leung (1986), the CH$_3^+$ reacts with
methanol, and after dissociative recombination leads to dimethyl
ether by
\begin{equation}
CH_3^+ + CH_3OH \rightarrow CH_3OCH_4^+ + h\nu
\stackrel{e^-}{\rightarrow} CH_3OCH_3 + H
\end{equation}

These works point out the importance of the ionic species in the
increase of interstellar molecular complexity. In star-forming
regions many ions could be produced by the photodissociation of
large organic molecules. Therefore, the knowledge of the
photodissociation processes and its ion yields plays an essential
role in interstellar chemistry.

The absence of more doubly ionized fragments in the PEPICO spectra
indicates that the doubly ionized acetic acid dissociates
preferentially via charge separation. The dynamics of the doubly
and triple ionized acetic acid molecule will be studied in a
future publication.

From Table~\ref{tab-path1} we can see that two of the main
photodissociation pathways of acetic acid occurs with the rupture
of the C-C bond, releasing CH$_3$$^+$ and COOH$^+$ ions, however
the yield of COOH$^+$ (12.8\%) undergoes little enhancement with
respect to CH$_3^+$ (11.6\%). This small excess could be
associated with the high stabilization of COOH$^+$ due to charge
resonance (charge migration) (Silverstein \& Webster 1998) or a
possible dissociation of CH$_3^+$ into minor fragments. Other
possibility is that the methyl carbon IP occurs at 291.6, 5 eV
less than for the carboxyl carbon (see Robin et al. 1988). As a
consequence, the number of resonances to excited orbitals
involving the carbon atom in the COOH site is higher than for the
other carbon site at photon energies above the IP of the methyl
carbon. This scenario could lead to a small preference for the
carboxyl carbon during photoelectron excitation/ionization at
these energies (and energies somewhat higher) which, after
dissociation, retain the charge.

\begin{table}[!t]
\caption{\textbf{Main photodissociation pathways from single
ionization}} \label{tab-path1}
\begin{center}
\setlength{\tabcolsep}{4pt}
\begin{tabular}{|l c l|}
\hline
$CH_3COOH + h\nu$ & $\longrightarrow$ & $CH_3COOH^+  + e^-$    \\
\hline \hline
$CH_3COOH^+$    & $\stackrel{14.1\%}{\longrightarrow}$ & $CH_3CO^+ + OH$ or ($O + H$) \\
                & $\stackrel{13.1\%}{\longrightarrow}$ & $H^+ + $ neutrals \\
                & $\stackrel{12.8\%}{\longrightarrow}$ & $COOH^+ + CH_3$ or ($CH_2 + H$) \\
                & $\stackrel{11.6\%}{\longrightarrow}$ & $CH_3^+ + COOH$ or ($CO_2 + H$; $CO + OH$) \\
                & $\stackrel{7.1\%}{\longrightarrow}$ & $CH_2^+ + HCOOH$ or ($H + COOH$) \\
                & $\stackrel{6.7\%}{\longrightarrow}$ & $COH^+ + CH_3 + O $ \\
                & $\stackrel{4.2\%}{\longrightarrow}$ & $CH^+ + $ neutrals \\
                & $\stackrel{3.5\%}{\longrightarrow}$ & $CH_2CO^+ + H_2O$ or ($OH + H$) \\
                & $\stackrel{2.7\%}{\longrightarrow}$ & $CO^+ + CH_3 + OH$ \\
                & $\stackrel{2.7\%}{\longrightarrow}$ & $O^+ +$ neutrals \\
                & $\stackrel{2.7\%}{\longrightarrow}$ & $OH^+ + CH_3CO$ or ($CH_3 + CO$) \\
                & $\stackrel{1.5\%}{\longrightarrow}$ & $CHCO^+ + $ neutrals \\
                & $\stackrel{1.4\%}{\longrightarrow}$ & $CO_2^+ + H + CH_3 $ or ($CH_2 + H$) \\

\hline
\end{tabular}
\end{center}
\end{table}

\subsection{Absolute photoionization and photodissociation cross sections}

The absolute cross section values for both photoionization and
photodissociation processes of organic molecules are extremely
important as the input for molecular abundances models (Sorrell
2001). In these theoretical model biomolecules are formed inside
the bulk of icy grain mantles photoprocessed by starlight
(ultraviolet and soft X-rays photons). Its main chemistry route
was based on radical-radical reactions followed by chemical
explosion of the processed mantle that ejects organic dust into
the ambient gaseous medium. For example, the number density  of a
given biomolecule in a steady state regime of creation and
destruction inside a gaseous-dusty cloud is given by
\begin{equation}
N_{Mol} = \frac{\dot{N}_{Mol} n_d}{<\sigma_{ph-d}> I_{0}}
\end{equation}
were $\dot{N}_{Mol}$ is the molecule ejection rate which depends
mainly on the molecule mass, the properties of the grains and the
cloud (see eq. 21 of Sorrell 2001). $n_d$ is the dust space
density, $I_{0}$ is the flux of ionizing photons (photons
cm$^{-2}$ s$^{-1}$) inside the cloud and $<\sigma_{ph-d}>$ is the
average photodissociation cross section in the wavelength range of
the photon flux density.

As mentioned by Sorrell (2001), the main uncertainty in this
equilibrium abundance model comes from $\sigma_{ph-d}$. Therefore
the precise determination of $\sigma_{ph-d}$ of biomolecules is
very important to estimate the molecular abundance of those
molecules in the interstellar medium. Moreover, knowing the photon
dose $\phi$ and $\sigma_{ph-d}$ values its is possible to
determine the half-life of a given molecule, as discussed by
Bernstein et al. (2004).

The photodissociation rates, $R$, of a molecule dissociated by the
interstellar radiation field $I_\varepsilon$ in the energy range
$\varepsilon_2 - \varepsilon_1$ is given by
\begin{equation} \label{eq-R}
R = \int_{\varepsilon_1 }^{\varepsilon_2} \sigma_{ph-d}(\varepsilon)
I_0(\varepsilon) d\varepsilon
\end{equation}
where $\sigma_{ph-d}(\varepsilon)$ is the photodissociation cross
section as a function of photon energy (cm$^2$), $I_0(\varepsilon)$
is the photon flux as a function of energy (photons cm$^{-2} eV^{-1}
s^{-1}$) (see discussion in Cottin et al. 2003 and Lee 1984).

\begin{figure}[!tb]
\resizebox{\hsize}{!}{\includegraphics{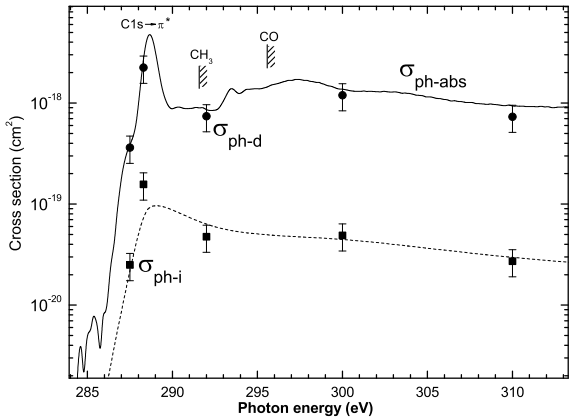}}
\caption{Non-dissociative single ionization (photoionization) cross
section ($\sigma_{ph-i}$) and dissociative ionization
(photodissociation) cross section ($\sigma_{ph-d}$) of acetic acid
as a function of photon energy. The photoabsorption cross-section
($\sigma_{ph-abs}$) taken from Robin et al. (1988) is also shown.}
 \label{fig-sigma}
\end{figure}

The half-life may be obtained from Eq.~\ref{eq-R} by writing
$t_{1/2}=ln 2 / R$, which does not depend on the molecular number
density.

In order to put our data on an absolute scale, after subtraction
of the linear background and false coincidences coming from
aborted double and triple ionization (see Simon et al 1991), we
have summed the contributions of all cationic fragments detected
and normalized them to the photoabsorption cross sections measured
by Robin et al (1988).

Assuming a negligible fluorescence yield (due to the low carbon
atomic number (Chen et al 1981)) and anionic fragments production
in the present photon energy range, we adopted that all absorbed
photons lead to cationic ionizing process. Therefore the
non-dissociative single ionization (photoionization) cross section
$\sigma_{ph-i}$ and the dissociative single ionization
(photodissociation) cross section $\sigma_{ph-d}$ of acetic acid
can be determined by
\begin{equation}
\sigma_{ph-i} = \sigma^{+} \frac{PIY_{CH_3COOH^+}}{100}
\end{equation}
and
\begin{equation}
\sigma_{ph-d} = \sigma^{+} \Big( 1 - \frac{PIY_{CH_3COOH^+}}{100}
\Big)
\end{equation}
where $\sigma^{+}$ is the cross section for single ionized fragments
(see description in Boechat-Roberty et al. 2005).

Both cross sections can be seen in Figure~\ref{fig-sigma} as a
function of photon energy. The absolute absorption cross section
of acetic acid (Robin et al. 1988) is also shown for comparison.
Those values are also shown in Table~\ref{tab-sigma}.

\begin{table}[!htb]
\centering \caption{Values of non-dissociative single ionization
(photoionization) cross section ($\sigma_{ph-i}$) and dissociative
ionization (photodissociation) cross section ($\sigma_{ph-d}$) of
acetic acid as a function of photon energy. The estimated total
error is 30\%. The photoabsorption cross section
($\sigma_{ph-abs}$) from Robin et al. (1988) is also shown.}
\label{tab-sigma}
\begin{tabular}{ l l c c c }
\hline \hline
Photon       &  & \multicolumn{3}{c}{Cross Sections (cm$^{2}$)}\\
\cline{3-5} energy (eV)  &  & $\sigma_{ph-d}$ & $\sigma_{ph-i}$     & $\sigma_{ph-abs}$ \\
\hline
285          &  & 3.61 $\times 10^{-19}$  & 2.49 $\times 10^{-20}$  & 4.12 $\times 10^{-19}$ \\
288.3        &  & 2.23 $\times 10^{-18}$  & 1.56 $\times 10^{-19}$  & 2.58 $\times 10^{-18}$ \\
292          &  & 7.40 $\times 10^{-19}$  & 4.75  $\times 10^{-20}$ & 8.52 $\times 10^{-19}$ \\
300          &  & 1.19 $\times 10^{-18}$  & 4.89 $\times 10^{-20}$  & 1.37 $\times 10^{-18}$ \\
310          &  & 7.32 $\times 10^{-18}$  & 2.72 $\times 10^{-20}$  & 9.31 $\times 10^{-19}$ \\
\hline \hline
\end{tabular}
\end{table}

\section{Summary and conclusions}

The goal of this work was to experimentally study ionization and
photodissociation processes of a glycine precursor molecule,
CH$_3$COOH (acetic acid). The measurements were taken at the
Brazilian Synchrotron Light Laboratory (LNLS), employing soft X-ray
photons from a toroidal grating monochromator (TGM) beamline (100 -
310 eV). The experimental set-up consists of a high vacuum chamber
with a time-of-flight mass spectrometer (TOF-MS). Mass spectra were
obtained using coincidence techniques.

We have shown that X-ray photon interactions with acetic acid
release a considerable number of energetic fragments, some of them
with high kinetic energy (ex. H$^+$, H$_2^+$ and OH$^+$). Unlike
the previous work with formic acid performed in the same spectral
range (Boechat-Roberty et al. 2005), no very high kinetic energy
fragments have been observed due the single photoionization of
acetic acid.

Several ionic fragments released from acetic acid
photodissociation have considerable kinetic energy. An extension
of this scenario to interstellar medium conditions suggests the
possibility endothermic ion-molecule (or radical-molecule)
reactions and this becomes important in elucidating the pathways
of formation of complex molecules (Largo et al. 2004).

Dissociative and non-dissociative photoionization cross sections
were also determined. We found that about 4-6\% of CH$_3$COOH
survive the soft X-ray ionization field. CH$_3$CO$^+$ and COOH$^+$
were the main fragments produced by high energy photons. The
former may indicate that the production-destruction cycle of
acetic acid in hot molecular cores could decrease the H$_2$O
abundance, since the net result of this process is the conversion
of water into OH + H. The latter ion plays an important role in
ion-molecule reactions to form large biomolecules like glycine.

%
\begin{acknowledgements} The authors would like to thank the staff
of the Brazilian Synchrotron Facility (LNLS) for their valuable
help during the experiments. We are particularly grateful to Dr.
R. L. Cavasso and Professor A. N. de Brito for the use of the
Time-of-Flight Mass Spectrometer. This work was supported by LNLS,
CNPq and FAPERJ.
\end{acknowledgements}
%
%

\end{document}